\documentclass[10pt,letterpaper]{article}
\usepackage{times}
\usepackage{epsfig}
\usepackage{multirow}
\usepackage{sectsty}
\usepackage{cite}
\usepackage{psfrag}
\usepackage{color}
\usepackage{amssymb}
\usepackage{marvosym}
\usepackage{amsbsy}
\usepackage{amsmath}
\usepackage{booktabs}
\usepackage{bm}
\usepackage{tabto}
\usepackage{graphicx}
\usepackage{subcaption}
\usepackage{floatrow}
\usepackage{scalerel}
\usepackage[colorlinks,bookmarksopen,bookmarksnumbered,citecolor=black,urlcolor=red]{hyperref}
\usepackage[labelfont=bf]{caption}

\usepackage{draftwatermark}
\SetWatermarkText{Draft}
\SetWatermarkScale{1.5}

\sectionfont{\normalsize \bfseries}
\subsectionfont{\normalsize \bfseries}
\begin{document}
\thispagestyle{empty}
\pagestyle{empty}
 \vspace{\baselineskip}
 \vspace{\baselineskip}
\begin{center}
	\Large{\textbf{On Dynamic Substructuring of Systems with Localised Nonlinearities}}\\
	\normalsize
	\vspace{\baselineskip}
	\vspace{\baselineskip}
	\vspace{\baselineskip}
	\vspace{\baselineskip}
	\large Thomas Simpson$^{1}$, Dimitrios Giagopoulos$^{2}$, Vasilis Dertimanis$^{1}$and Eleni Chatzi$^{1}$\\ 
	\vspace{\baselineskip}
	$^{1}$ Institute of Structural Engineering, Department of  Civil, Environmental and Geomatic Engineering, ETH Zürich, Stefano-Franscini Platz 5, 8093, Zürich, Switzerland\\
	$^{2}$ Department of Mechanical Engineering, University of Western Macedonia,\\ Kozani 50100, Greece\\
\end{center}
\normalsize
\vspace{\baselineskip}
\vspace{\baselineskip}
\vspace{\baselineskip}

\section*{ABSTRACT}
\label{abstract}
Dynamic substructuring methods encompass a range of techniques, which allow for the decomposing of large structural systems into multiple coupled subsystems. This decomposition of structures into smaller domains has the principle benefit of reducing computational time for dynamic simulation of the system by considering multiple smaller problems rather than a single global one. In this context, dynamic substructuring methods may form an essential component of hybrid simulation, wherein they can be used to couple physical and numerical substructures at reduced computational cost. Since most engineered systems are inherently nonlinear in nature, particular potential lies in incorporating nonlinear treatment in existing sub-structring schemes, which are mostly developed within a linear problem scope. \\

The most widely used and studied dynamic substructuring methods are classical linear techniques such as the Craig-Bampton and MacNeal-Rubin methods. These methods are widely studied and implemented in many commercial FE packages. However, as linear methods they naturally break down in the presence of nonlinearities. Recent advancements in substructuring have involved the development of enrichments to the linear substructuring methods, which allow for some degree of nonlinearity to be captured. The use of mode shape derivatives has been shown to be able to capture geometrically non-linear effects as an extension to the Craig-Bampton method. Other candidates include the method of Finite Element Tearing and Interconnecting. Linear substructuring methods have been used in several cases for hybrid simulation, rendering additional benefits in removing non-physical high frequency modes which may be excited due to controller tracking errors in hybrid simulations.\\

In this work, a virtual hybrid simulation is presented in which a linear elastic vehicle frame supported on four nonlinear spring damper isolators is decomposed into separate domains. One domain consisting of the finite element model of the vehicle frame, which is reduced using the Craig-Bampton method to only include modes relevant to the structural response. The second domain consists of the nonlinear isolators whose restoring forces are characterised by nonlinear spring and damper forces. Coupling between the models is carried out using a Lagrange multiplier method and time series simulations of the system are conducted and compared to the full global system with regards to simulation time and accuracy.\\

\textbf{Keywords: Dynamic Substructuring, Hybrid Simulation, Nonlinearity} 

\section{INTRODUCTION}
In this study we investigate implementation of dynamic substructuring schemes with hybrid simulation. Dynamic substructuring (DS) methods are used to decompose dynamic models into several separate regimes or substructures. The global model of the system is always recoverable via use of coupling constraints enforced between substructures. The key benefit of DS methods lies in the capability of analyzing each substructure separately and possibly in parallel, eventually assembling the global solution; this can greatly reduce the computational burden of a given analysis. However, further benefits are generated from the implementation of different analysis schemes on different regions of a structure. A key motivation of the above being the ability to individually treat nonlinear regions, without the requirement for adoption of a global nonlinear solution method \cite{Voormeeren2011}.\\

Dynamic substructuring methods are typically combined with reduction methods, such as Component Mode Synthesis (CMS), whereby reduced order models of the individual substructures are determined, which further decreases the computational load of the solution process. Most traditional techniques such as the Craig-Bampton and Rubin-Macneal methods \cite{VanderValk2010}, rely on a modal decomposition of the substructures whereby only a subset of mode shapes is retained -those relevant to the structural response- whilst the remaining modes are ignored. The commonly adopted linear schemes are naturally not suitable for the treatment of nonlinear systems. Recent advancements in substructuring have involved the development of enrichments to the linear substructuring methods, \cite{Wu2018}, \cite{Kim2015}, \cite{Kim2017}, which allow for some degree of nonlinearity to be captured. The use of mode shape derivatives has been shown to capture geometrically non-linear effects as an extension to the Craig-Bampton method \cite{Wu2018}.\\

An alternative to these methods is the so-called finite element tearing and interconnecting (FETI) technique \cite{Farhat1991} in which a coupling method making use of Lagrange multipliers is used, whilst the reduction of the substructures can be carried out separately. This allows for the use of any generic meta-modelling technique for the reduction of the non-linear substructures, such as a non-linear autoregressive models or a Gaussian process regression. 
These techniques can be shown to provide accurate representation of a complex model over a given input range of interest \cite{ChatziCostas}, \cite{Mai2016}, \cite{inproceedings}. The FETI method can also allow for the coupling of nonlinear substructures or linear to nonlinear substructures. This has previously been used in hybrid simulations \cite{Abbiati2019} and was shown to perform well.\\

 Hybrid testing, which forms a focus of the European ITN Project DyVirt, on Dynamic Virtualization, involves the setup of coupled dynamic tests in which a numerical system in a computer is coupled to a physical component in a dynamic testing rig. The system is then solved as a whole by integrating the numerical system, calculating predicted forces at the interface between the structures and applying these to the physical system. The response of the physical system is then measured and fed back into the equations of the numerical system \cite{BensonShing2008}. The benefit of hybrid testing is that simple or well characterised components of a system can be represented numerically, whilst more complex or critical components can be physically tested. This allows for a reduction in cost and complexity of experimentation against full scale testing with an increase in fidelity and reduction in uncertainty with respect to purely numerical modelling. The coupling between the physical and numerical components within hybrid testing is a parallel of the DS problem, which attempts to couple numerical systems; as such, similar methods to those in DS can be used \cite{Wagg}.\\

 Within the domain of hybrid testing the``gold standard" is considered to be real time hybrid testing . Real time testing implies that the time scales of the numerical and physical system are the same; and that this time scale matches that of ``wall time". The benefit of real time testing is that it allows for rate dependent behaviour and nonlinearities of the physical system to be accounted for \cite{BensonShing2008}. However as we attempt to approach real time testing the problem of the time required to integrate the numerical substructure becomes an issue. In order to carry out a real time hybrid simulation, each step of the numerical integration must be able to be carried out within a smaller time increment than the integration time step. This becomes a problem when numerical substructures of increasing complexity are considered and leaves the options of either increasing the time step of the integration which can have negative effects on accuracy and stability \cite{Pegon2008}, or finding a reduced order model of the numerical system which can be solved more quickly. However, there remains considerable potential for the use of model order reduction techniques in combination with hybrid simulation. \\
 
 Application of CMS methods in hybrid testing have been made as in \cite{Abbiati2019}, although in this case the key purpose of the dynamic substructuring was to remove high frequency modes from the numerical system to yield improved stability characteristics, rather than to improve the speed of calculation.\\

Another issue which arises in hybrid simulation is that as the integration time step is increased to allow for real time simulation, the time step for control of the physical system is also increased to match it in order to couple the systems. This increased time step of the physical system can result in a non-smooth operation of the actuator due to control commands being sent at such a large interval. This can cause issues with regards to the fidelity of the applied loads and issues of stability in the solution \cite{doi:10.1002/eqe.743}. As such, when considering the solution time step not only must the numerical integration be considered but rather also the physical control system. This problem has inspired algorithms which allow for adoption of individual time steps in the physical and numerical subsystems \cite{Abbiati2018}, \cite{Abbiati2019a}.

\section{Theory}
The key elements used in the virtual hybrid simulation we present herein  are the model order reduction carried out on the linear "numerical" system, and the integration algorithm which solves the systems and imposes coupling between them.

\subsection{Craig-Bampton Reduction}\label{Craig-Bampton Reduction}

The Craig-Bampton is one of the most prominent and long standing techniques in CMS and is already implemented in numerous commercial finite element softwares \cite{Craig1968}. 
In the Craig-Bampton method each sub-structure can be reduced individually and then usually the reduced sub-structures are coupled together using a primal assembly. However, in this case only the linear system will be reduced whilst the coupling procedure will be carried out within the integration algorithm.\\

The key aspects of the reduction are that the degrees of freedom (DOFs) of each substructure are partitioned into internal DOFs, $x_i$, and external DOFs, $x_b$, i.e., those which lie on the interface between substructures.
\begin{equation}
    \begin{bmatrix}
    M_{ii} & 
    M_{ib} \\
    M_{bi} & 
    M_{bb}
    \end{bmatrix}
    \begin{bmatrix}
    \ddot{x_{i}} \\ \ddot{x_b}
    \end{bmatrix} +
    \begin{bmatrix}
    K_{ii}&K_{ib}\\
    K_{bi}&K_{bb}
    \end{bmatrix}
    \begin{bmatrix}
    x_{i}\\x_b
    \end{bmatrix}=0
\end{equation}
The internal DOFs, which in most cases are the majority, can be significantly reduced using a modal decomposition from the eigenvalue problem in Equation \ref{eq:inMod}. Given that for structural dynamics the lowest eigenmodes are dominant, the majority of these modes can be discarded, with only the lower modes, $\Phi_r,$ retained. These are known as the fixed interface normal modes.\\

\begin{equation}
    M_{ii}^{-1}K_{ii}=\Phi\Sigma\Phi^T,\quad \Phi=\begin{bmatrix}
    \Phi_r&\Phi_d\end{bmatrix}
\label{eq:inMod}
\end{equation}
 
In addition to the fixed interface normal modes, constraint modes, $\Psi$, are also used to characterise the boundary DOFs and the static response. Constraint modes correspond to the static deformation shape due to a unit displacement applied at a boundary DOF. As such there exist as many, as there exist boundary DOFs. These are calculated as follows.
\begin{equation}\label{eq:constraintmodes}
    \Psi=-K_{ii}^{-1}K_{ib}
\end{equation}

Using both the fixed interface and constraint modes, the original high dimensional coordinate set can be approximated on a significantly lower coordinate set, with the reduction of DOFs dependant upon the number of fixed interface modes that are discarded. The reduced coordinate set consists of the generalised internal coordinates $q$ and the retained boundary coordinates $x_b$.

\begin{equation}
\begin{bmatrix}
x_i\\x_b
\end{bmatrix}
\approx
\begin{bmatrix}
    \Phi_r & \Psi
    \\0& I
\end{bmatrix}
\begin{bmatrix}
q\\x_b
\end{bmatrix}
=CB
\begin{bmatrix}
q\\x_b
\end{bmatrix}
\end{equation}
The mass and stiffness matrices of the sub-structures can then be reduced by projection onto these reduced coordinate sets. 
\begin{equation}
    \hat{K}=CB^TKCB
    \quad   
    \hat{M}=CB^TMCB
\end{equation}
In the Craig-Bampton method the reduction basis consists of the generalised internal coordinates, whilst the boundary coordinates are retained as physical; this simplifies the coupling procedure.

\subsection{Integration and Coupling}
The coupled integration scheme used herein is inspired by that presented in \cite{Abbiati2019a}, this comprises a technique designed for hybrid simulation in which each of the substructures is solved as a free solution over the time interval with coupling enforced at the end of the time interval using a method similar to the FETI method \cite{Farhat1991}. The partitioned design of this integration method allows it to be applied in a hybrid test in that, the 2 systems are solved in parallel with coupling only enforced at the end of each time step. This means that in the context of a hybrid test, displacements can be enforced on the physical system, with the resulting restoring force being measured and being used to solve the free problem in the physical system and finally coupling being enforced at the end of the time step.\\

The coupled equations of motion for the numerical and physical systems to be integrated are shown in equations \ref{eq:Num} and \ref{eq:Phys}. These present the typical equations of motion of a nonlinear mechanical system in state space form. However, these equations are augmented with the term $L\Lambda$, this represents the coupling forces between the two systems with $\Lambda$ being the Lagrange multipliers giving the interface force intensities and $L$ a boolean matrix which locates the interface forces on the overall system. 

\begin{equation}
\label{eq:Num}
M^N\dot{Y}^{N}_{n+1}+R^N(Y^N_{n+1})=L^N\Lambda_{n+1}+F^N_{n+1}
\end{equation}

\begin{equation}
M^P\dot{Y}^{P}_{n+1}+R^P(Y^N_{n+1})=L^P\Lambda_{n+1}+F^P_{n+1}
\label{eq:Phys}
\end{equation}

Each of the physical and numerical subsystems is solved over a given time interval. This is initially carried out as a free problem, i.e., ignoring the coupling of the other substructures within this interval. For the free solution of the substructures a trapezoidal integration rule was used which corresponds to a $\gamma$ parameter =0.5. The trapezium rule integration uses a prediction and a correction step in order to step forward in the integration. The prediction step as indicated in equation \ref{eq:fpred} gives the predicted state at the next integration step based on the previous state and state rate values, the integration parameter $\gamma$ and the time step $\Delta T$.

\begin{equation}
    \Tilde{Y}_{n+1}^{F}=Y_n+(1-\gamma)\Delta T\dot{Y}_n
    \label{eq:fpred}
\end{equation}
Within equation \ref{eq:fpred}, $\Tilde{Y}_{n+1}^{F}$ indicates the predicted state of the free solution at the next time step which will be corrected to find the final free solution $Y_{n+1}^F$. The Predicted state is made based on the solutions from the previous coupled time step: $Y_n$ and $\dot{Y}_n$.\\

The predicted state rate is then calculated using the predicted state, according to equation \ref{eq:fpredrate} making use of the matrix $D$ assembled as in equation \ref{eq:D}

\begin{equation}
    \dot{Y}_{n+1}^F=D^{-1}(F_{n+1}-R(\Tilde{Y}_{n+1}^F))
    \label{eq:fpredrate}
\end{equation}

\begin{equation}
    D=M+\gamma\Delta TR_0;
    \label{eq:D}
\end{equation}

In the above equations $F$ is the external force vector, $R(Y)$ evaluates the potentially nonlinear restoring force vector, consisting of the elastic and damping forces, evaluated at state $Y$ and D is a matrix as formed in equation \ref{eq:D} wherein $M$ is the mass matrix,  and $R_0$ the tangential restoring force vector linearised at zero state. This tangential restoring force vector was found using the ADIGATOR toolbox for automatic differentiation \cite{Patterson2013Aeo}. This toolbox allows the tangential stiffness matrix to be found numerically inclusive of systems with nonlinear restoring forces.\\

The correction step is then made as in equation \ref{eq:fcorr} making use of the predicted state and state rates found previously.

\begin{equation}
    Y_{n+1}^F=\Tilde{Y}_{n+1}^F+\gamma\Delta T\dot{Y}_{n+1}^F
    \label{eq:fcorr}
\end{equation}

Having found the free solution for each of the substructures individually and in parallel, the coupling can then be enforced between them. The coupling is enforced using a dual assembly procedure, by enforcing equal and opposite forcing at the boundary DOFs using Lagrange multipliers. Firstly, a matrix is assembled, as in equation \ref{eq:H}, which forms the so called Steklov-Poincar\'e operator \cite{Quarteroni}. This relates the discrepancy in state at the boundary nodes to the Lagrange multiplier intensities.

\begin{equation}
    H=G^ND_N^{-1}L^N+G^PD_P^{-1}L^P
    \label{eq:H}
\end{equation}
In equation \ref{eq:H} the matrices $G^N$ and $G^P$ are boolean matrices, which enforce compatibility of the substructures \cite{VanderValk2010} for the numerical and physical systems respectively, whilst $L^N$ and $L^P$ are boolean matrices, which locate the interface forces for the numerical and physical structures \cite{VanderValk2010}.\\

Having derived the Steklov-Poincar\'e operator, which is assembled only once at the beginning of the integration procedure, the Lagrange multipliers can be identified based on the free solutions of each of the substructures as in equation \ref{eq:lagrange}.

\begin{equation}
    \Lambda_{n+1}=-H^{-1}(G^NY_{n+1}^{N,f}+G^PY_{n+1}^{P,f})
    \label{eq:lagrange}
\end{equation}

The link solutions for each of the substructures are then found, these represent the effect on the state of the substructure of the interface forces due to coupling. These link solutions are found by first finding the state rate due to the interface forces as in equation \ref{eq:linkrate} and then the link state using this state rate as in equation \ref{eq:linkstate}.

\begin{equation}
    \dot{Y}^{P,L}_{n+1}=D_P^{-1}L^P\Lambda_{n+1} \quad \dot{Y}^{N,L}_{n+1}=D_N^{-1}L^N\Lambda_{n+1}
    \label{eq:linkrate}
\end{equation}

\begin{equation}
    Y^{P,L}_{n+1}= \gamma\Delta T\dot{Y}^{P,L}_{n+1} \quad Y^{N,L}_{n+1}= \gamma\Delta T\dot{Y}^{N,L}_{n+1}
    \label{eq:linkstate}
\end{equation}

Finally the global solution for each of the substructures taking into account the coupling between them is found as in equation \ref{eq:Global} by summing the free and link solutions for each of the substructures.

\begin{equation}
    Y^P_{n+1}=Y^{P,F}_{n+1}+Y^{P,L}_{n+1} \quad Y^N_{n+1}=Y^{N,F}_{n+1}+Y^{N,L}_{n+1}
    \label{eq:Global}
\end{equation}

\subsection{With Sub-cycling}
In order to ameliorate the problem of actuator smoothness in hybrid simulations, wherein the numerical integration time step is necessarily larger than the control frequency of the actuator, a set of methods exist in which the physical system is controlled at a smaller time step than the numerical system, with coupling enforced at the larger numerical time step \cite{Abbiati2019a}. In connection with the above described algorithm, the sub-cycling method can be implemented in the calculation of the free solution of the physical substructure.\\

Firstly, a number of sub-cycles is decided upon, defined as the ratio of the numerical time step to the physical time step. The integration scheme with sub-cycling operates similarly as described previously except that the free solution for the physical substructure, as described by equations \ref{eq:fpred}-\ref{eq:fcorr} above, is now replaced by the following procedure.\\

The sub-cycling loop runs over the number of sub-cycles $ss$. It is initiated by the coupled solution at the previous numerical system time step. The loop ends by finding the free solution at the next numerical time step. Equation \ref{eq:fpredss} shows the prediction step of the trapezium rule in which $\Delta T^P$ is the time step of the physical system.

\begin{equation}
    \Tilde{Y}_{n+\frac{j}{ss}}^{P,F}=Y_{n+\frac{j-1}{ss}}^{P,F}+(1-\gamma)\Delta T^P \dot{Y}_{n+\frac{j-1}{ss}}^{P,F}
    \label{eq:fpredss}
\end{equation}

Equation \ref{eq:fpredratess} calculates the state rate prediction in which it is worth noting that the Lagrange multipliers from the last coupling time step are now included in the solution and that the D matrix is formed as in equation \ref{eq:D} albeit with the finer time step of the physical system used in the assembly.

\begin{equation}
    \dot{Y}_{n+\frac{j}{ss}}^{P,F}=D_P^{-1}(F^P_{n+\frac{j}{ss}}-R(\Tilde{Y}_{n+\frac{j}{ss}}^{P,F})+L^P\lambda(1-\frac{j}{ss}))
    \label{eq:fpredratess}
\end{equation}

Finally, the corrected free solution is calculated as in equation \ref{eq:fcorrss}.

\begin{equation}
    Y^{P,F}_{n+\frac{j}{ss}}=\Tilde{Y}_{n+\frac{j}{ss}}^{P,F}+\gamma \Delta T^P\dot{Y}_{n+\frac{j}{ss}}^{P,F}
    \label{eq:fcorrss}
\end{equation}

The sub-cycling loop then ends after $ss$ sub-cycles whereby the $n+1$ prediction of the free state of the physical system is found and the coupling procedure can then be followed as described in the previous section.

\section{CASE STUDY}

\begin{figure}[h!]
    \centering
    \includegraphics{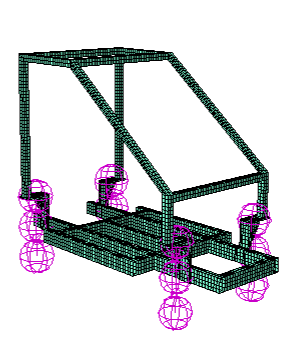}
    \caption{Vehicle frame structure model with attachment points of suspension indicated.}
    \label{fig:fullframe}
\end{figure}

The case study considered herein is a steel vehicle-like frame considered to be linear elastic, which is mounted on four nonlinear suspension structures. Computer models of the frame and suspension structures were created based on physical specimens. The frame structure model, along with the mounting points are presented in figure \ref{fig:fullframe}.\\

The frame structure is considered to be linear elastic and is meshed using a combination of quadrilateral shell elements and hexahedral solid elements with a total of 45,564 DOFs. The material properties of the frame are a Youngs modulus of $210\,GPa$, a Poissons' ratio of 0.3 and a density of $7800\,kgm^{-3}$. This model has previously been calibrated to a physical structure, details of which along with dimensions of the frame can be seen in \cite{Giagopulos2006}. The nonlinear suspension systems consist of a base excited mass representing the wheel and its interaction with the road surface, alongside a linear spring and nonlinear damping element which connect the wheel mass with the frame structure. These nonlinear suspension elements have also been identified from a physical component connected to the physical frame in a previous study \cite{Tsotalou2017}. The best model found to approximate the non linearity in the damper was found to be a dry friction type damping force \cite{Giagopulos2006}. The restoring force equations of the nonlinear suspension system are shown in equation \ref{eq:restforce}, in which $k_1$ represents the linear stiffness coefficient, $c_1$ the linear damping coefficient and, $c_2$ and $c_3$ the nonlinear damping coefficients. The coefficients are summarised below.

\begin{equation}
    f_{r}(x)=k_{1}x \quad \quad f_{d}({\dot{x}})=c_{1}\dot{x}+\frac{c_2\dot{x}}{c_3+\left|\dot{x}\right|}
    \label{eq:restforce}
\end{equation}

\begin{equation}
   m=0.160, \,\, k_1=35, \,\, c_1=0.65, \,\, c_2=10, \,\, c_3=0.55
\end{equation}

\begin{figure}[H]
    \centering
    \includegraphics[width=80mm]{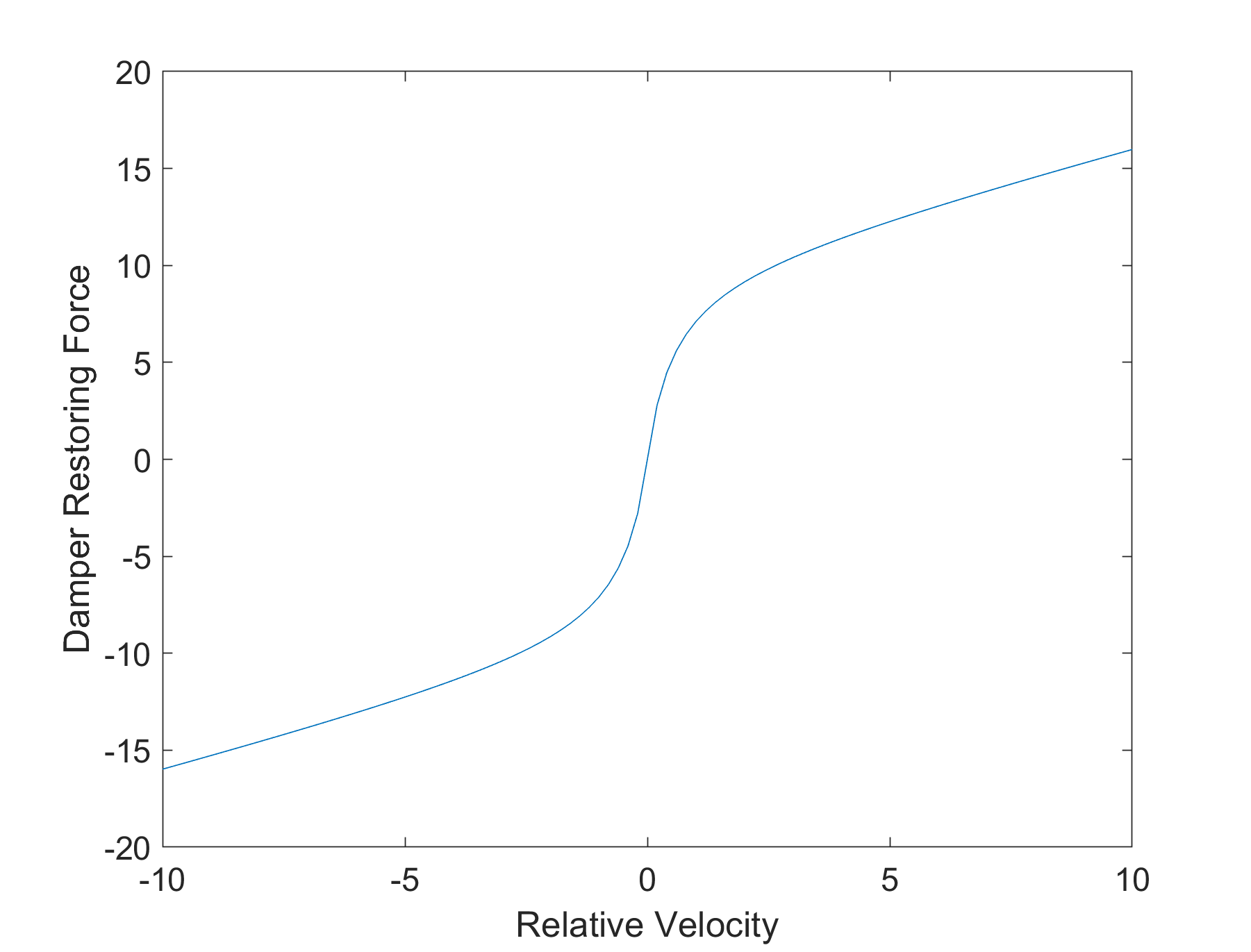}
    \caption{Restoring force curve of the nonlinear damper elements}
    \label{fig:nonlindamp}
\end{figure}

The form of the nonlinearity present in the damper is illustrated in figure \ref{fig:nonlindamp} where it is seen that the frictional nonlinearity delivers a more pronounced effect around the zero point, whilst further from zero more linear behaviour is observed.

\begin{figure}[h!]
    \centering
    \begin{subfigure}[b]{0.59\textwidth}
    \centering
        \includegraphics[width=70mm]{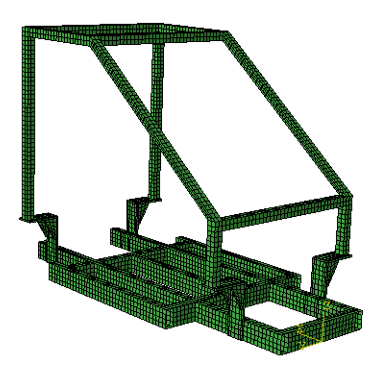}
        
        \caption{Linear FE mesh of the vehicle frame substructure}
       
    \end{subfigure}
    \begin{subfigure}[b]{0.39\textwidth}
    \centering
        \includegraphics[width=50mm]{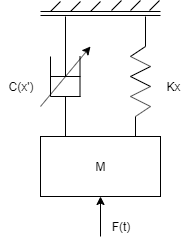}
      
        \caption{Nonlinear suspension subsystem model}
          \label{fig:tiger}
    \end{subfigure}
    \caption{Linear and nonlinear substructures of the vehicle frame and suspension}
    \label{fig:Nss and Pss}
\end{figure}

Figure \ref{fig:Nss and Pss} presents the linear FE mesh of the frame structure and a diagrammatic representation of the nonlinear suspension systems. In the case of this virtual hybrid simulation, the linear frame is considered to be the numerical system which is to be reduced in order, whilst the suspension systems are considered to be the physical systems to be solved in full order.

\section{Virtual Hybrid Simulation}

A virtual hybrid simulation was carried out in which the case study described above was considered. The linear frame structure was considered to be the numerical system with the four suspension substructures considered as the physical systems. The linear system was first reduced using a Craig-Bampton reduction, and then the partitioned integration procedure was implemented on the substructures.

\subsection{Reduction of the Linear Frame}
The linear substructure was reduced according to the methodology described in section \ref{Craig-Bampton Reduction}. As described earlier, a subset of fixed interface mode shapes were retained, which represent the dynamic behaviour of the structure within the frequency region of interest. The reduction basis was formed on the 30 lowest fixed interface normal mode shapes of the frame, along with 4 constraint modes each corresponding to the vertical DOFs at the 4 mounting points of the suspension systems.\\

\begin{figure}[h!]
    \centering
    \begin{subfigure}[b]{0.49\textwidth}
    \centering
        \includegraphics[width=85mm]{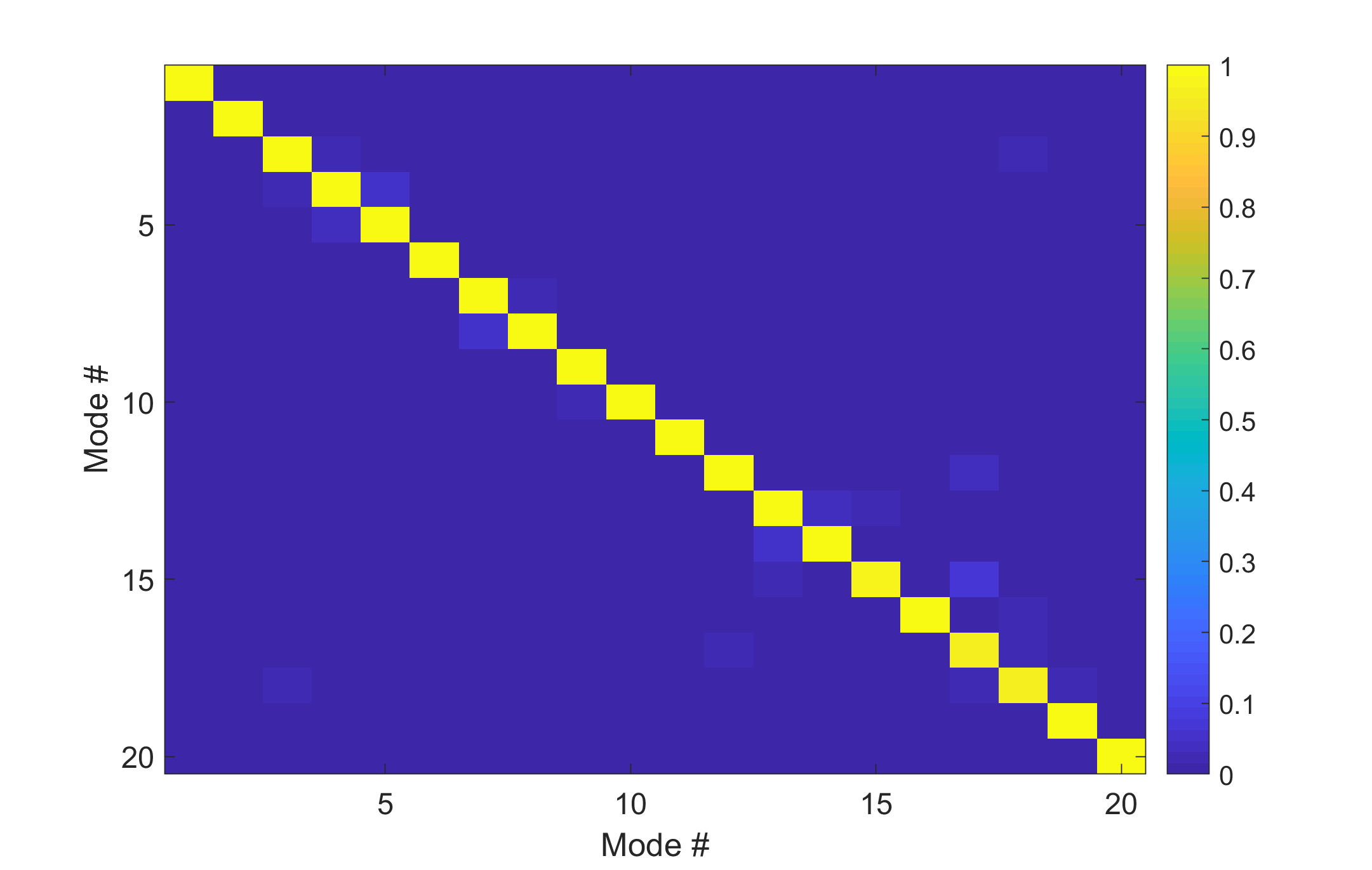}
        \caption{}
        \label{fig:eigvecs}
    \end{subfigure}
    \begin{subfigure}[b]{0.49\textwidth}
    \centering
        \includegraphics[width=70mm]{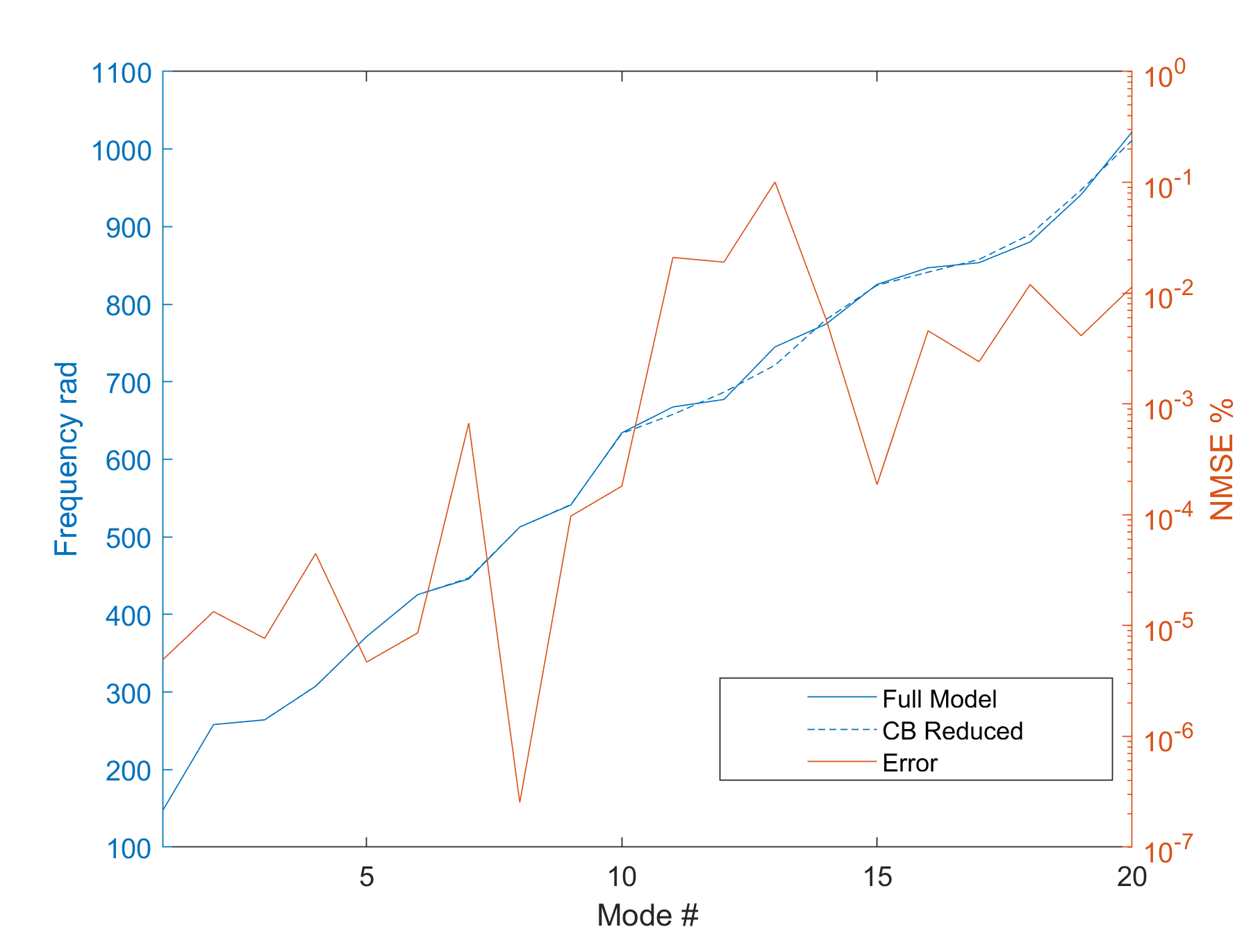}
        \caption{}
        \label{fig:eigvals}
    \end{subfigure}
    
    \caption{Mode shape and natural frequency fidelity of the CB reduced system}
    \label{fig:eigvecsvals}
\end{figure}

\begin{figure}[H]
    \centering
    \begin{subfigure}[b]{0.49\textwidth}
    \centering
        \includegraphics[width=80mm]{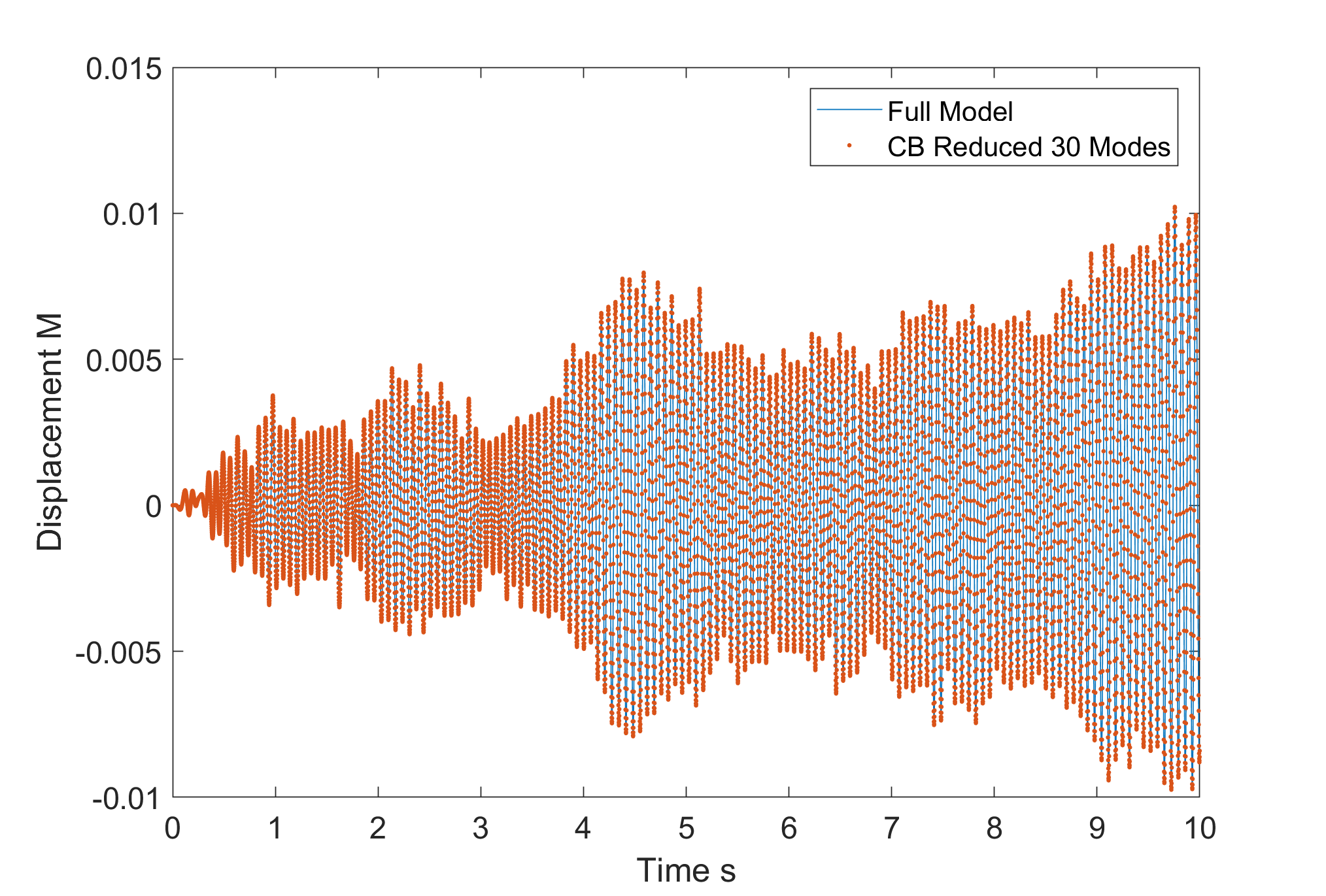}
        \caption{}
        
    \end{subfigure}
    \begin{subfigure}[b]{0.49\textwidth}
    \centering
        \includegraphics[width=80mm]{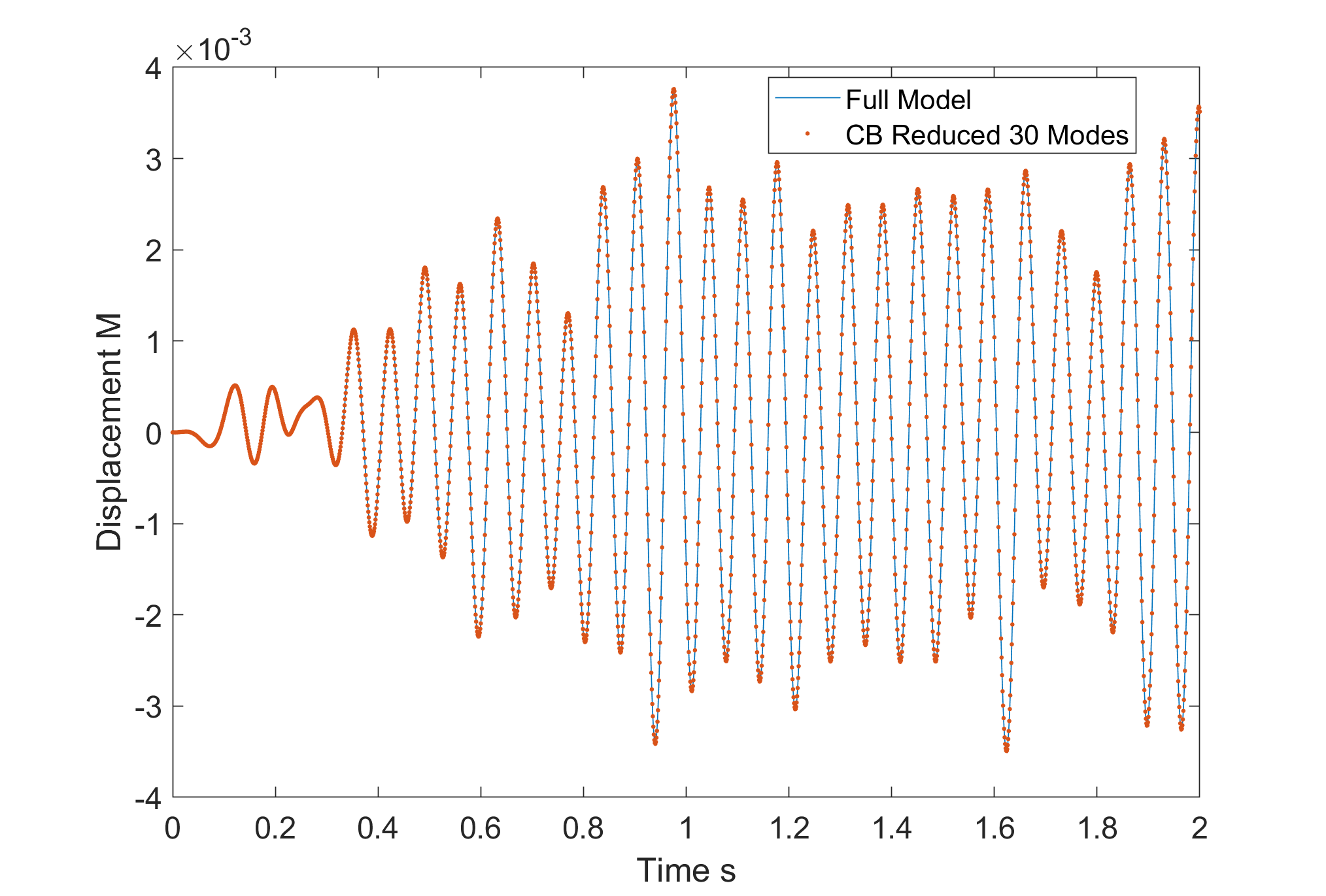}
        \caption{}
    \end{subfigure}
    
    \caption{Time series response fidelity of the CB reduced system}
    \label{fig:CBtimesim}
\end{figure}

Figures \ref{fig:eigvecsvals} and \ref{fig:CBtimesim} demonstrate the fidelity of the reduction basis relative to the full model of the frame structure. In figure \ref{fig:eigvecsvals} a comparison is made between the 20 lowest natural frequencies and modal shapes between the full model and the CB reduction. Figure \ref{fig:eigvecs} shows the cross modal assurance criterion \cite{Pastor2012} between the mass orthogonalised mode shapes of the full system and the CB reduced system. This gives a metric for the orthogonality of the mode shapes to one another and is calculated as in equation \ref{eq:MAC}.
\begin{equation}
   MAC(\Phi_{CB},\Phi_{F})=\frac{
   \begin{vmatrix}
   \Phi_{CB}^T\Phi_F
   \end{vmatrix}^2}{(\Phi_{CB}^T\Phi_{CB})(\Phi_F^T\Phi_F)}
    \label{eq:MAC}
\end{equation}
For identical mode shapes the diagonal of the plot should show one whilst all off diagonal elements should be zero. The plot demonstrates the fidelity of the mode shapes, as all diagonal elements are very close to one whilst all off diagonals are close to zero. Figure \ref{fig:eigvals} compares the first 20 natural frequencies of the reduced and full systems, along with the normalised mean square error (NMSE) between them. The natural frequencies of the systems are very similar with none of the first 20 natural frequencies exhibiting an error larger than 0.1\%\\

Figure \ref{fig:CBtimesim} compares the time history response of both the reduced and full order systems to the same input. The input used in this case was a band limited white noise signal between 0 and 200 Hz with a variance of 1 which was applied in the vertical direction at the four nodes to which the suspension systems connect. This frequency region was chosen as the region of interest as ISO 8608, detailing the classification of road surface profiles, considers the region of interest to be up to a spatial frequency of $17.77\,rad/m$, at an assumed vehicle speed of $100\,km/h$. This corresponds to a maximum frequency of interest of $78.6\,Hz$ \cite{Mucka2017}. The frequency range of interest was then selected to be comfortably above this value so as to ensure fidelity within the region.\\

The reconstruction of the time series response is near perfect for this input with a mean square error value of $1.04e-08$. According to both the eigenvalue and time series analyses the Craig-Bampton reduced system shows excellent fidelity to the full order model when excited. In both cases a Newmark integration scheme was used in the time series simulation.

\subsection{Comparison of Monolithic and Partitioned Solutions}

\begin{figure}[H]
    \centering
    \includegraphics[width=90mm]{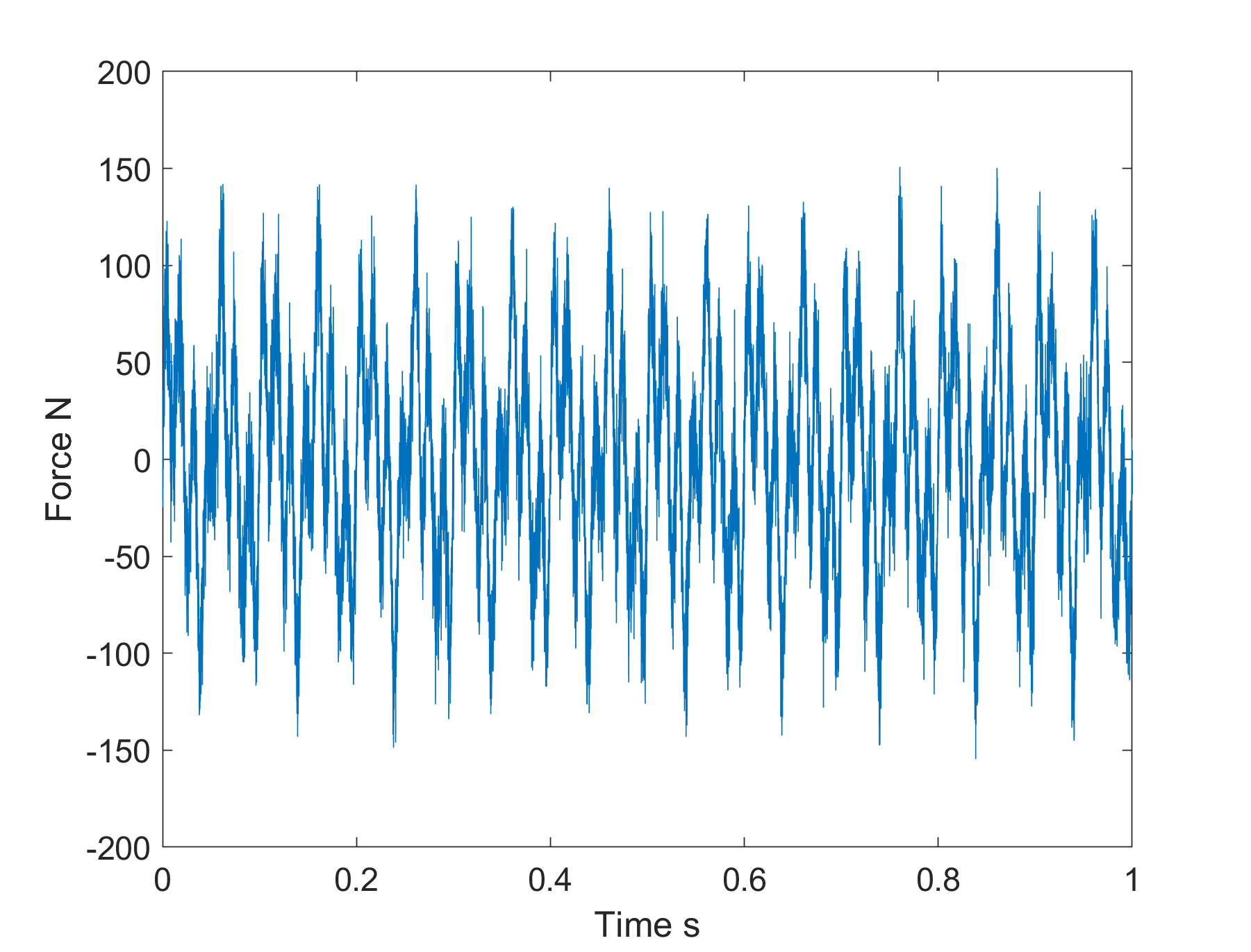}
    \caption{White noise corrupted multi sine input signal}
    \label{fig:AG}
\end{figure}

To assess the performance of the partitioned solution of the reduced order system, it was compared to a monolithic solution of the full system with regards to both fidelity and simulation time. The inputs chosen for the simulation were all multi sine excitations corrupted by white noise. The inputs were applied as base motion to the masses in the suspension substructure. Whilst each of the suspension substructures were excited with the same multi-sine content, they differed in the white noise corruption. Figure \ref{fig:AG} shows an exemplar signal applied to one of the four suspension systems.

\begin{figure}[H]
    \centering
    \begin{subfigure}[b]{0.49\textwidth}
    \centering
        \includegraphics[width=80mm]{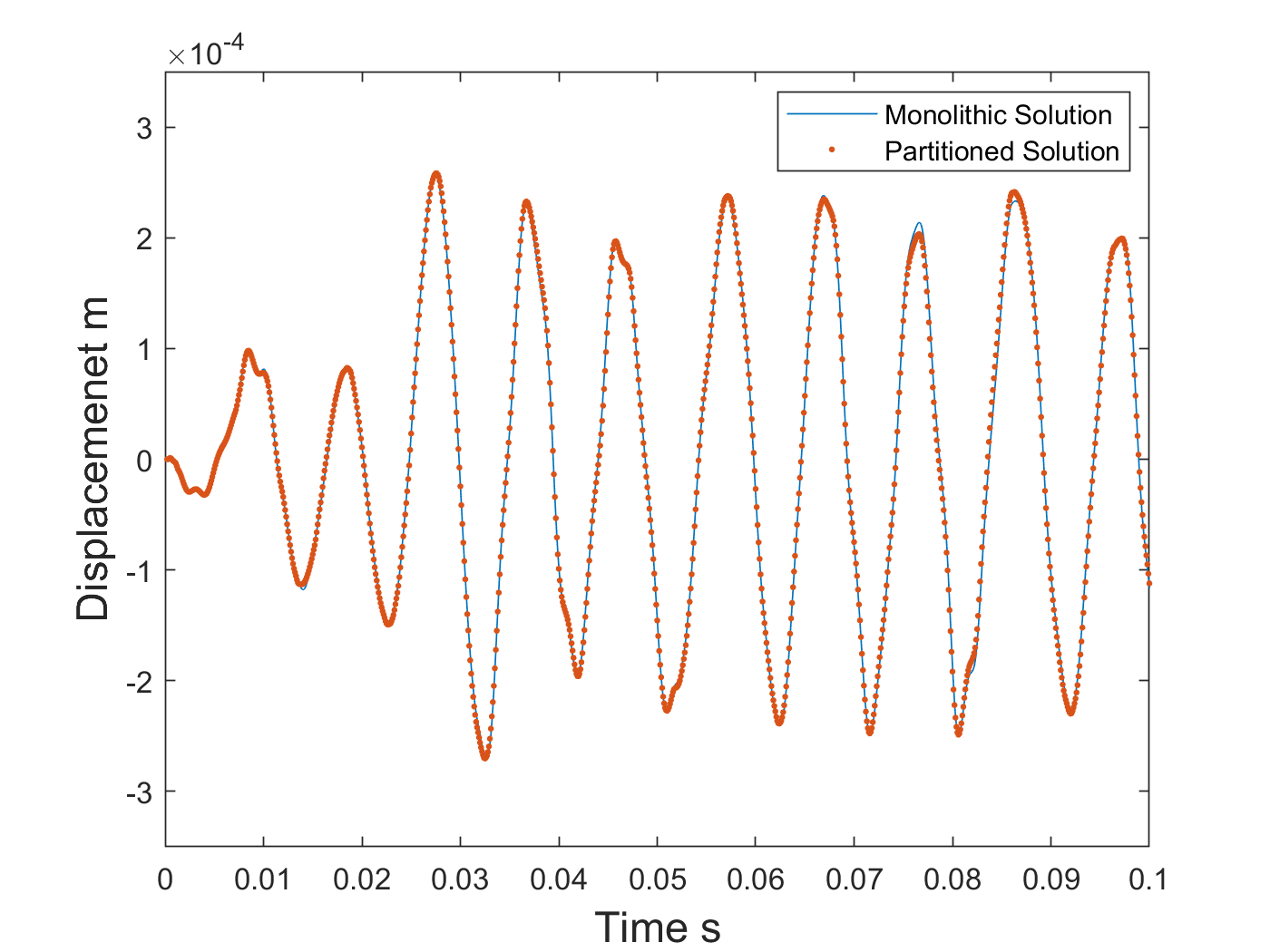}
        \caption{}
     
    \end{subfigure}
    \begin{subfigure}[b]{0.49\textwidth}
    \centering
        \includegraphics[width=80mm]{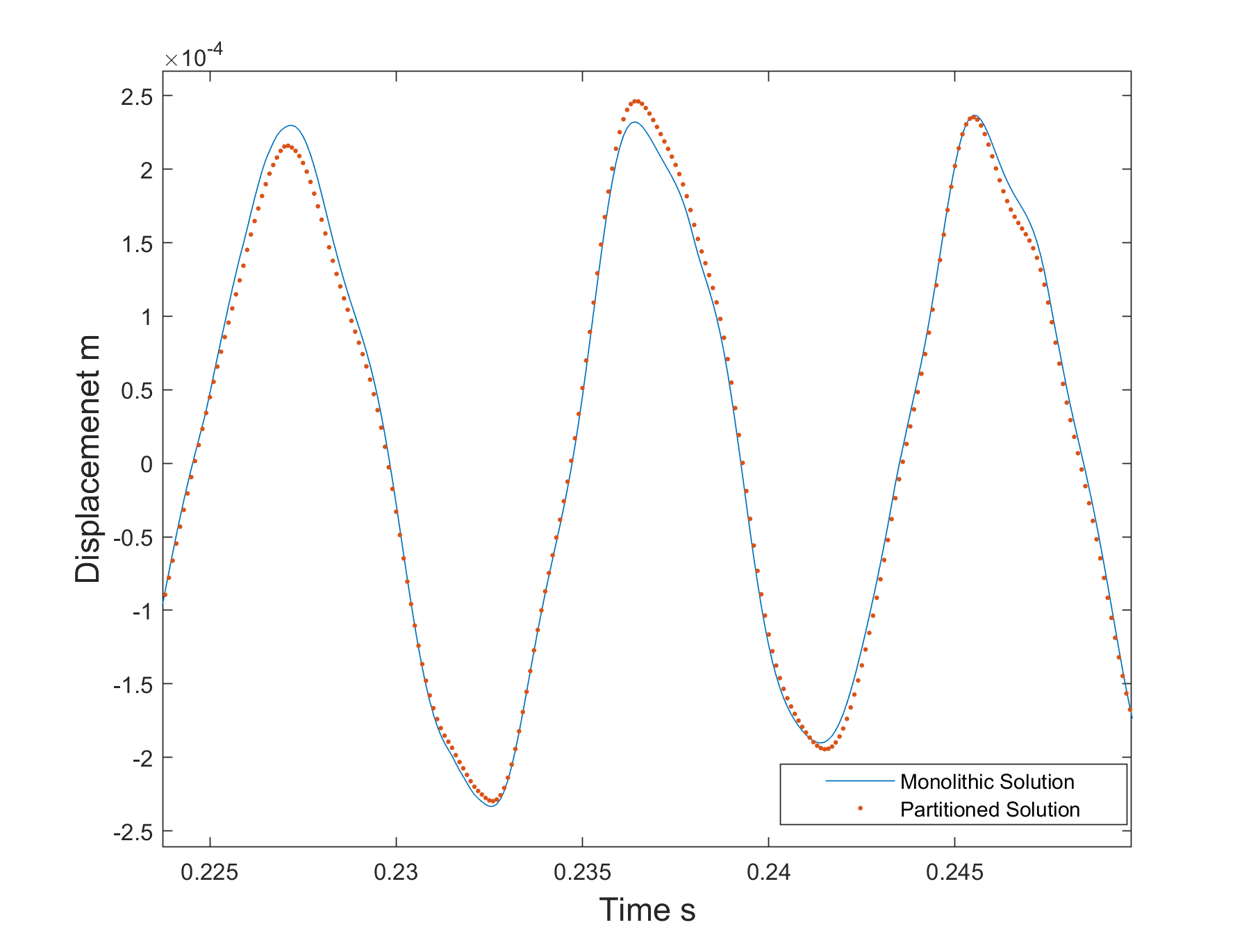}
        \caption{}
    \end{subfigure}
    
    \caption{Comparison of the monolithic and partitioned solution of the time series response of the reduced system}
    \label{fig:monopara}
\end{figure}

Figure \ref{fig:monopara} compares the response of the DS reduced system to the monolithic solution produced in Abaqus. The solutions show good agreement although some discrepancy can be seen particularly around the peaks. The mean squared error between the signals was $2.63e-11$ showing the good fidelity. Some discrepancy is expected between the responses firstly due to the reduced linear system not performing identically to the full linear system. Additionally, the tangential stiffness matrix which is used for the coupling procedure introduces errors as it is effectively a first order linearisation of the coupling process.\\

In both the monolithic and partitioned solutions displayed, a time step of $1e-3$ was used and 1 second of time history was simulated. This simulation from the monolithic solver required 1557 s compared to 0.15 s for the DS reduced system. This demonstrates the massive time savings possible using the reduction scheme. Additionally, it is noteworthy that the DS solution was found in faster than real time hence demonstrating its potential for real time hybrid simulation. It is worth noting that there are additional offline calculations required for the DS solution such as, the CB reduction and the forming of tangential stiffness matrices, the total offline time for these operations was 109 seconds. As such the overall time was still considerably less for the DS reduced system and furthermore the offline times are not of great concern with regards to hybrid simulation. \\

\begin{figure}[H]
    \centering
    \begin{subfigure}[b]{0.49\textwidth}
    \centering
        \includegraphics[width=80mm]{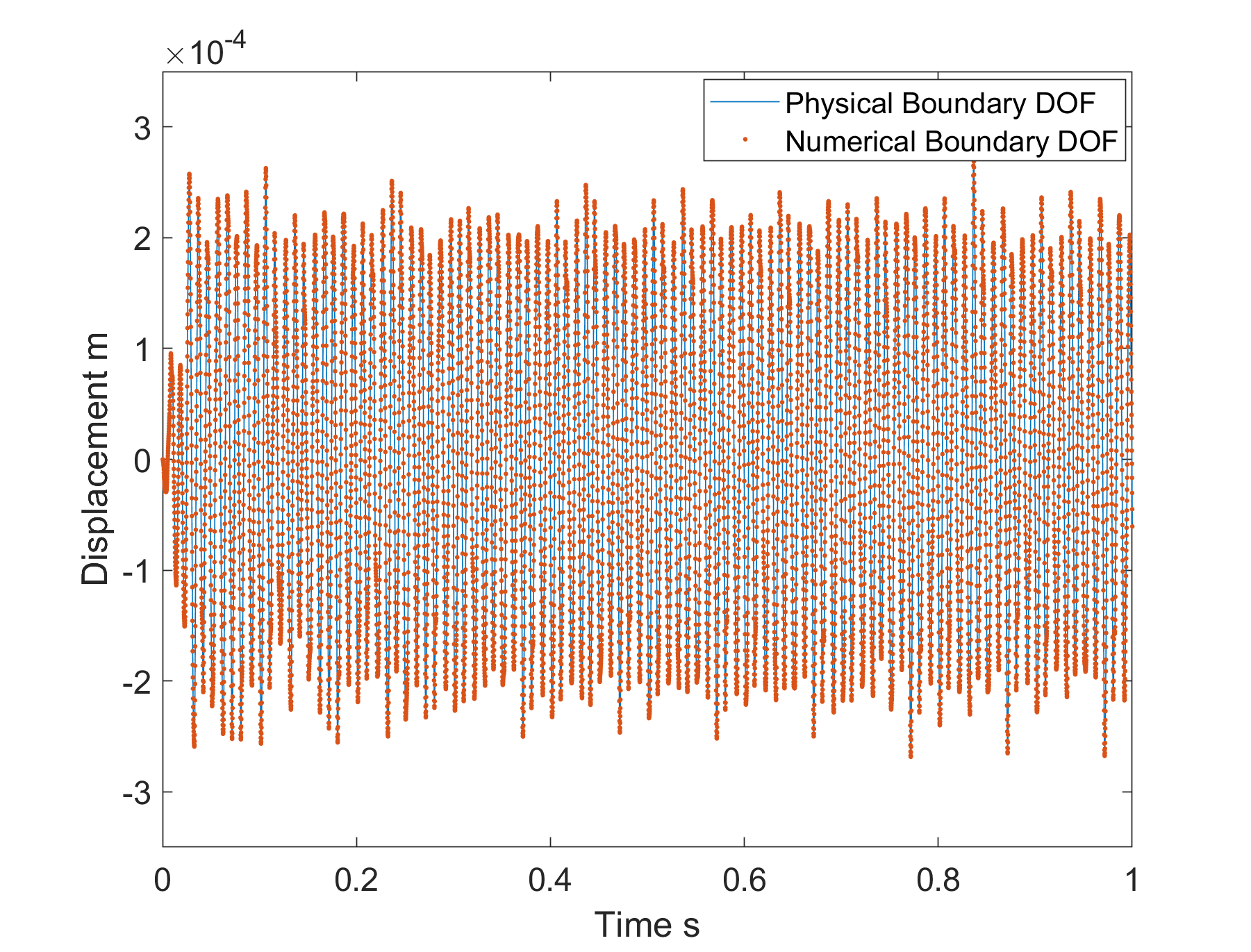}
        \caption{}
        
    \end{subfigure}
    \begin{subfigure}[b]{0.49\textwidth}
    \centering
        \includegraphics[width=80mm]{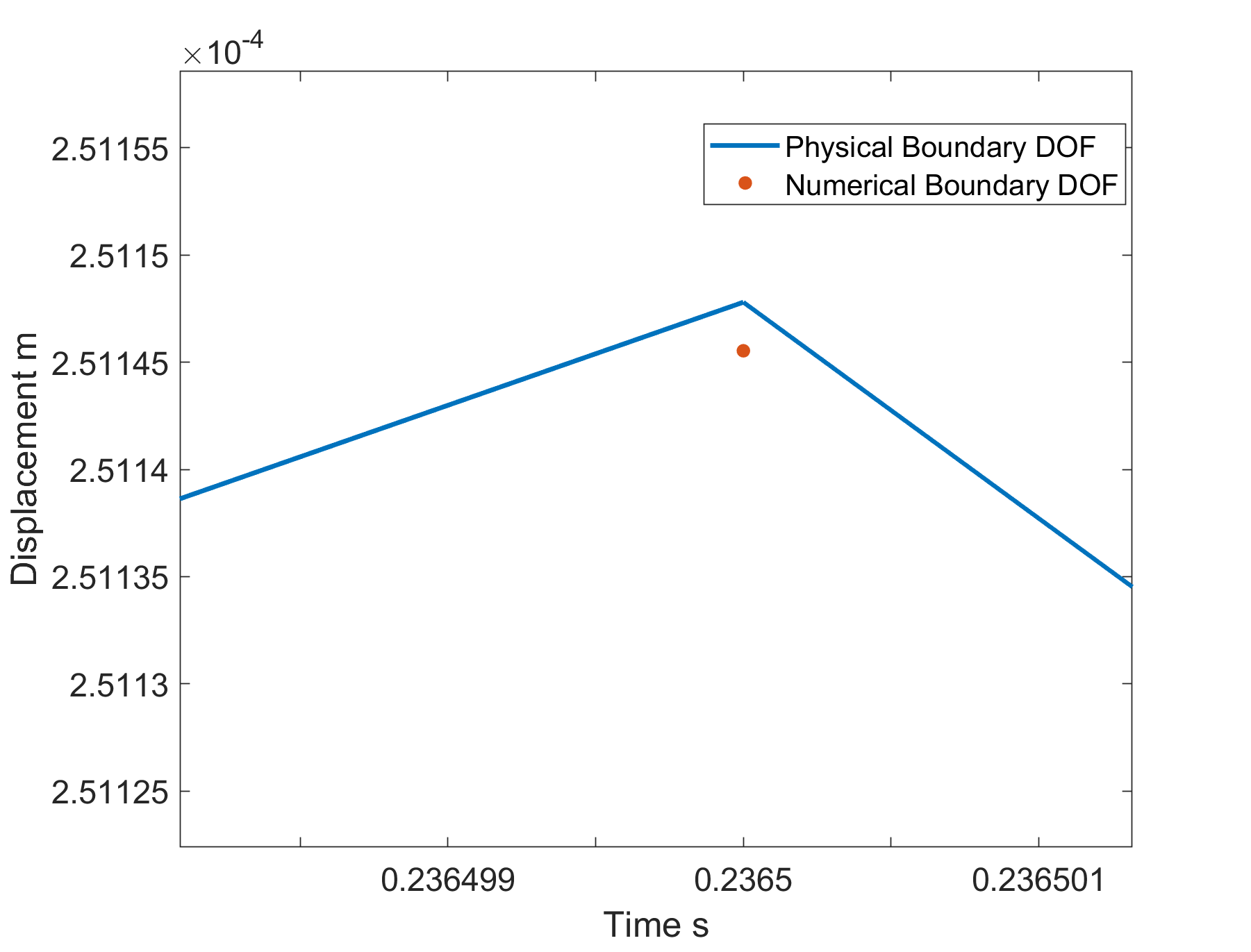}
        \caption{}
        \label{fig:linksolszoom}
    \end{subfigure}
    
    \caption{Response of physical and numerical
    boundary DOF}
    \label{fig:linksols}
\end{figure}

Figure \ref{fig:linksols} shows the solution of one of the boundary DOF of the numerical system and the boundary DOF of the physical system to which it is coupled. There is very close agreement between the two solutions showing that the coupling procedure was successful. However, figure \ref{fig:linksolszoom} shows that there is a slight discrepancy indicating that the coupling procedure does not enforce hard compatibility between the two substructures but rather that a slight discrepancy is allowed as a result of the dual assembly \cite{Rixen2004}.

\subsection{Subcycling}
\begin{figure}[H]
    \centering
    \begin{subfigure}[b]{0.49\textwidth}
    \centering
        \includegraphics[width=80mm]{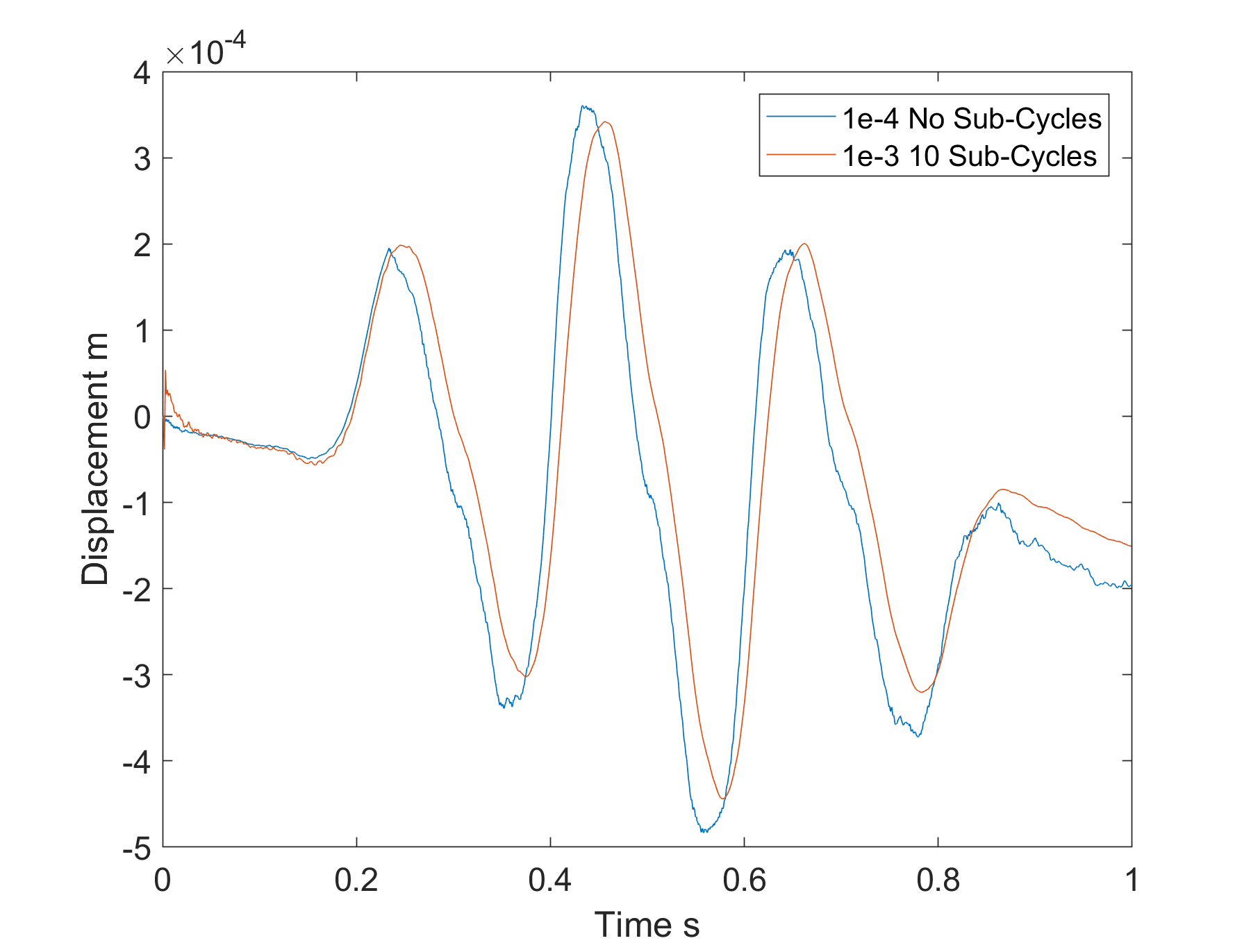}
        \caption{}
       
    \end{subfigure}
    \begin{subfigure}[b]{0.49\textwidth}
    \centering
        \includegraphics[width=80mm]{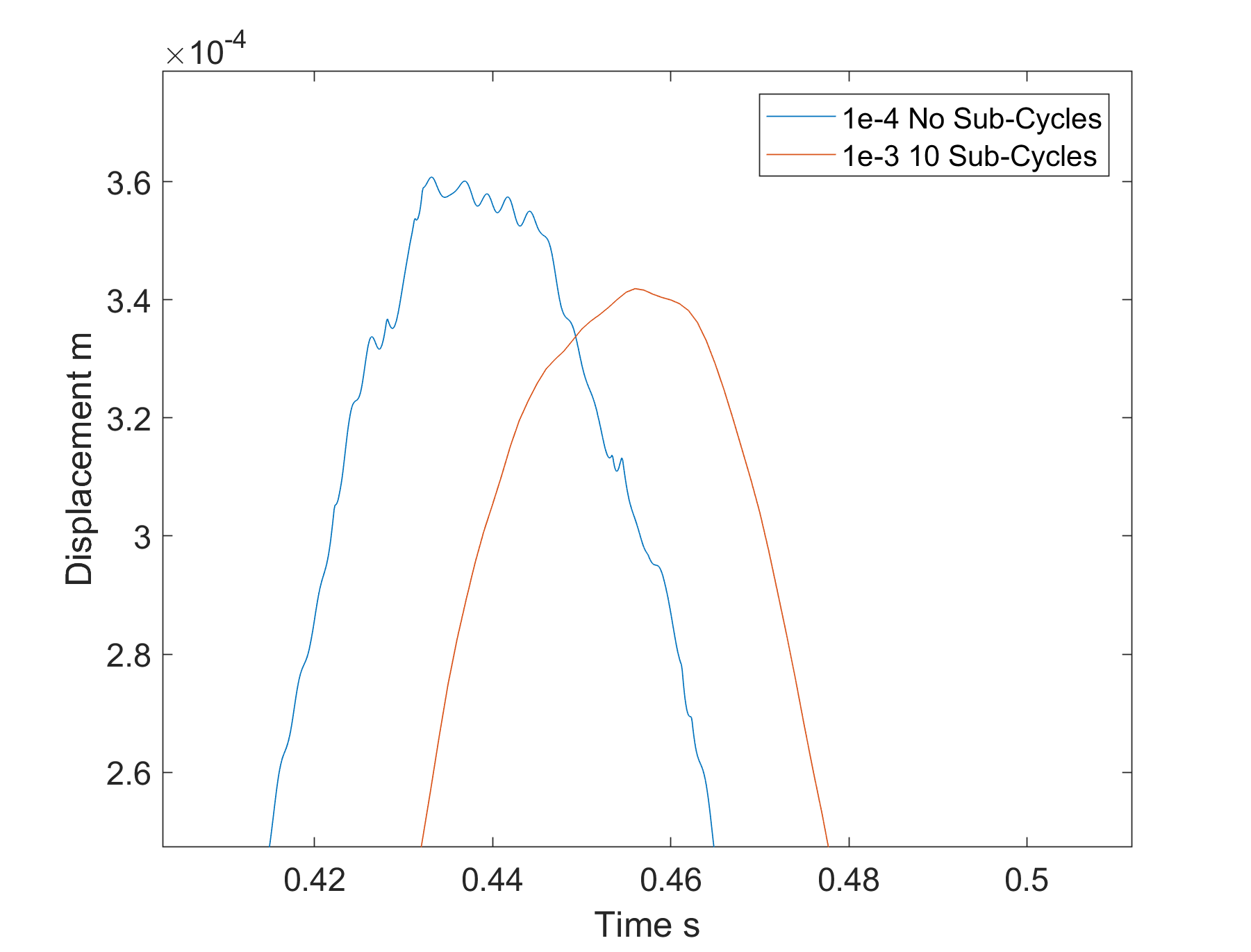}
        \caption{}
        \label{fig:withwithoutzoom}
    \end{subfigure}
    
    \caption{Response integrated at $\Delta T = 1e-4$ and $\Delta T=1e-3$ with 10 sub-cycles}
    \label{fig:withwithout}
\end{figure}

The effect of the addition of the sub-cycling of the physical system in the algorithm was also investigated. Firstly a comparison was made between the non sub-cycling algorithm with a time step of $1e-4$ and the sub-cycling algorithm with 10 sub-cycles and a time step of $1e-3$ this comparison is presented in figure \ref{fig:withwithout}. The overall response of the two algorithms are similar however, it is clear that there is some loss in fidelity. Overall, the sub-cycling algorithm exhibits stability issues. Previous implementations of the method in \cite{Abbiati2019a} and \cite{Abbiati2018} adopted an algorithm with numerical damping, as opposed to the undamped trapezium rule method used here. It is thought that the use of a damped algorithm could ameliorate these issues.\\ 

The algorithm shows potential however given that the sub-cycling algorithm can allow the time step of the physical system to be decreased by order 10 with an increase of computational time of approximately 2 times. It is also evident from figure \ref{fig:withwithoutzoom} that the sub-cycled system did exhibit a smoother response in the physical system, which should improve the control of the actuator. 


\begin{figure}[H]
    \centering
    \begin{subfigure}[b]{0.49\textwidth}
    \centering
        \includegraphics[width=80mm]{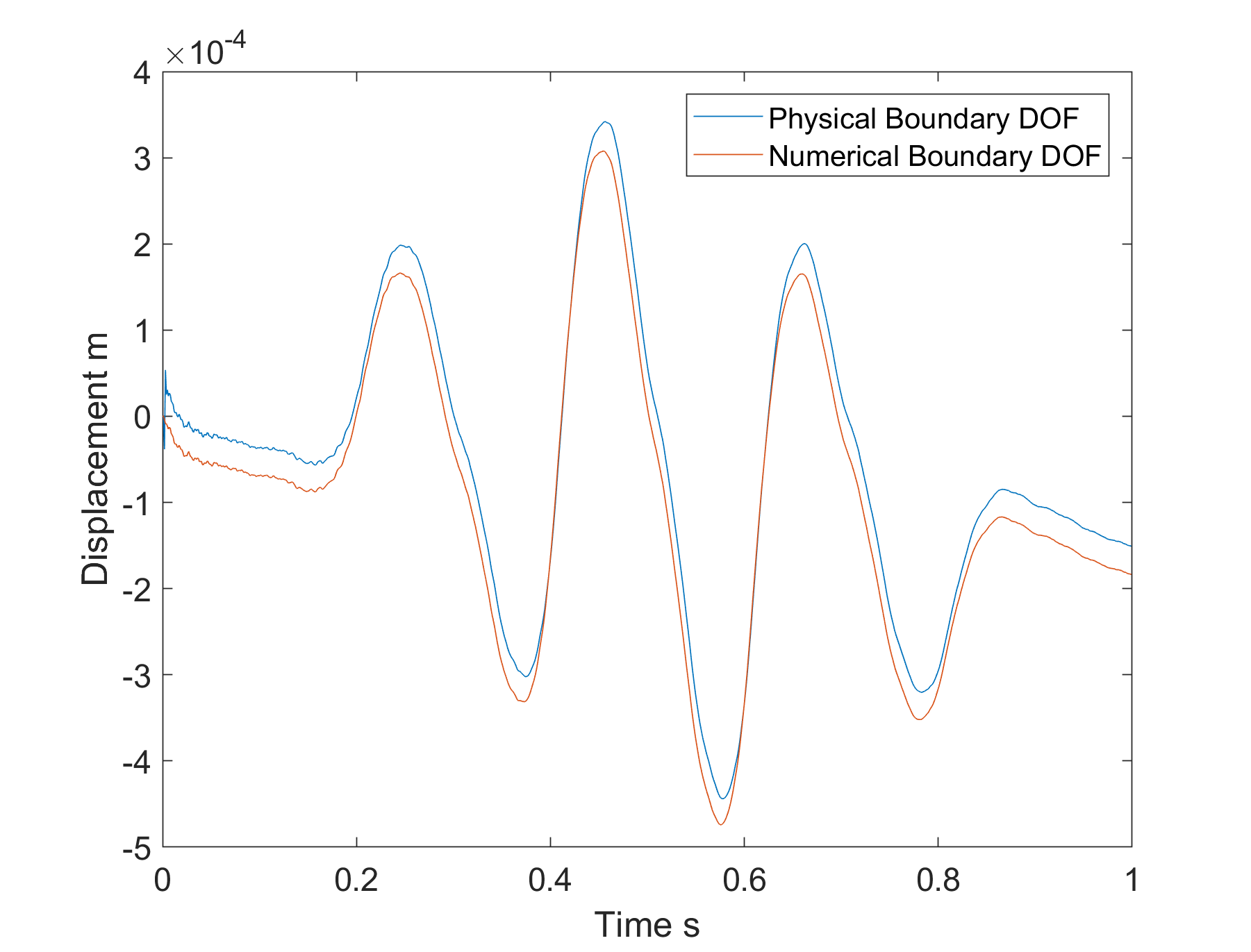}
        \caption{}
        
    \end{subfigure}
    \begin{subfigure}[b]{0.49\textwidth}
    \centering
        \includegraphics[width=80mm]{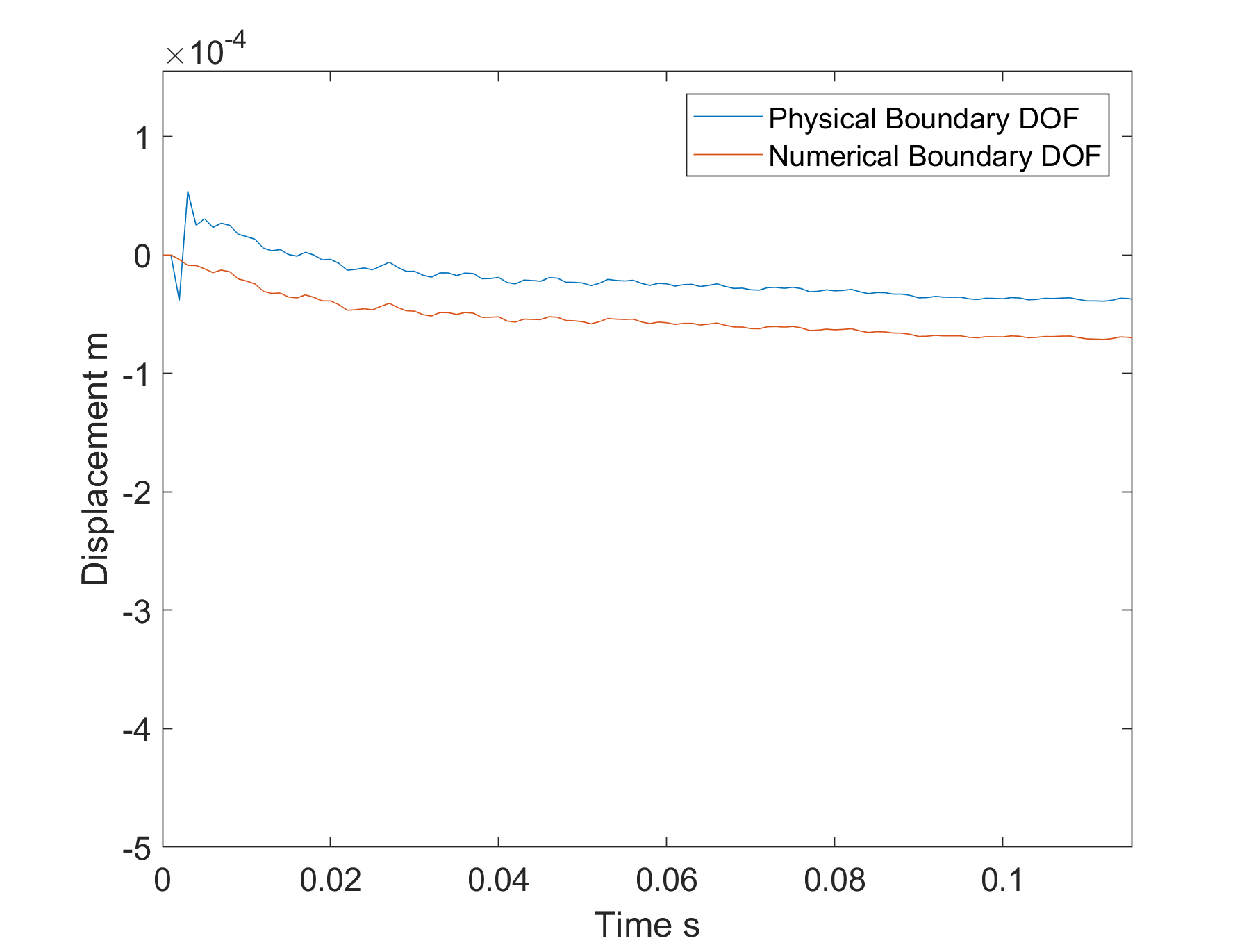}
        \caption{}
        \label{fig:numphysszoom}
    \end{subfigure}
    
    \caption{Response of physical and numerical boundary DOF with 10 sub-cycles}
    \label{fig:numphysss}
\end{figure}

Figure \ref{fig:numphysss} presents the response of a coupled boundary DOF on the numerical and physical substructures. Compared to the case without sub-cycling there is a clearly discrepancy, this discrepancy again highlights the capacity of the dual coupling procedure to allow for a soft coupling procedure. As seen in figure \ref{fig:numphysszoom} the response of the physical system is initially very uneven which seems to cause the discrepancy which is maintained throughout. This unevenness is once again thought to be a result of the instability in the sub-cycling procedure, as highlighted previously.

\section{CONCLUSIONS}
A dynamic sub-structuring process has been presented, which can be used to couple linear and nonlinear systems in a manner suitable to hybrid simulation. It was demonstrated that this method can considerably decrease the required computational time and was also shown to reduce the integration time sufficiently to implement a real time hybrid simulation. The reduced system was also shown to deliver good fidelity when compared to the full monolithic system. The dual assembly procedure carried out was shown to enforce coupling between the substructures without enforcing hard compatibility and allowed an amount of discrepancy.\\

The DS integration algorithm was also implemented using a sub-cycling procedure whereby the physical system was integrated at a time step smaller than that of the numerical system. This method had considerable problems with instability of the physical system but it is thought that with the use of an algorithm with numerical damping these problems could be improved. It was shown that the use of the sub-cycling method resulted in a smoother response in the physical system which has advantages for the control of the actuator related to the physical system.

\section{ACKNOWLEDGEMENTS}
\scalerel*{\includegraphics{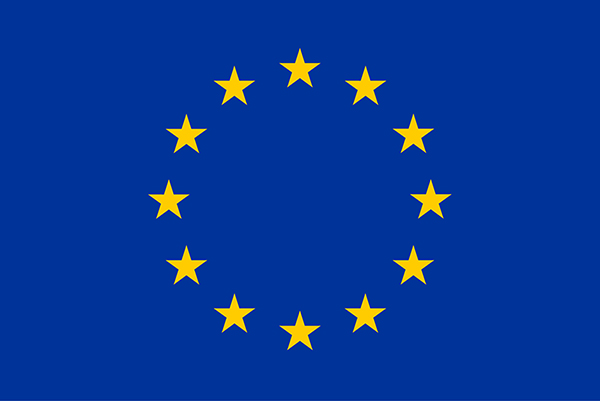}}{2*B} This project has received funding from the European Union’s Horizon 2020 research and innovation programme under the Marie Skłodowska-Curie grant agreement No 764547

\section*{REFERENCES}
\begingroup
\renewcommand{\section}[2]{}%
\bibliographystyle{unsrt}
\bibliography{bibmac}

\begin{thebibliography}{10}

\bibitem{Voormeeren2011}
S.~N. Voormeeren, P.~L.~C. van~der Valk, and D.~J. Rixen.
\newblock {Practical Aspects of Dynamic Substructuring in Wind Turbine
  Engineering}.
\newblock pages 163--185. Springer, New York, NY, 2011.

\bibitem{VanderValk2010}
P.L.C. van~der Valk.
\newblock {Model Reduction {\&} Interface Modeling in Dynamic Substructuring
  -Application to a multi-megawatt wind turbine-}.
\newblock page 177, 2010.

\bibitem{Wu2018}
L.~Wu.
\newblock {Model order reduction and substructuring methods for nonlinear
  structural dynamics}.
\newblock 2018.

\bibitem{Kim2015}
Jin-Gyun Kim and Phill-Seung Lee.
\newblock {An enhanced Craig-Bampton method}.
\newblock {\em International Journal for Numerical Methods in Engineering},
  103(2):79--93, jul 2015.

\bibitem{Kim2017}
Jeong-Ho Kim, Jaemin Kim, and Phill-Seung Lee.
\newblock {Improving the accuracy of the dual Craig-Bampton method}.
\newblock {\em Computers {\&} Structures}, 191:22--32, oct 2017.

\bibitem{Farhat1991}
Charbel Farhat.
\newblock {A method of finite element tearing and interconnecting and its
  parallel solution algorithm}.
\newblock Technical report, 1991.

\bibitem{ChatziCostas}
Eleni~N {Chatzi Costas} and Papadimitriou Editors.
\newblock {Identification Methods for Structural Health Monitoring}.

\bibitem{Mai2016}
C~V Mai, M~D Spiridonakos, E~N Chatzi, and B~Sudret.
\newblock {Surrogate modelling for stochastic dynamical systems by combining
  NARX models and polynomial chaos expansions}.
\newblock Technical report, 2016.

\bibitem{inproceedings}
Fabian Gruber, Johannes Rutzmoser, and Daniel Rixen.
\newblock {Comparison between primal and dual Craig-Bampton substructure
  reduction techniques}.
\newblock 2015.

\bibitem{Abbiati2019}
G~Abbiati, G~Miraglia, M~Petrovic, N~Mojsilovic, and B~Stojadinovic.
\newblock {Hybrid Simulation of the Seismic Response of a Masonry Facade Based
  on Component-Mode Synthesis}.
\newblock 2019.

\bibitem{BensonShing2008}
P.~{Benson Shing}.
\newblock {Real-Time Hybrid Testing Techniques}.
\newblock pages 259--292. Springer, Vienna, 2008.

\bibitem{Wagg}
D~J Wagg, P~Gardner, R~J Barthorpe, and K~Worden.
\newblock {On key technologies for realising digital twins for structural
  dynamics applications}.

\bibitem{Pegon2008}
P.~Pegon.
\newblock {\em Continuous PsD Testing With Substructuring}, pages 197--257.
\newblock Springer Vienna, Vienna, 2008.

\bibitem{doi:10.1002/eqe.743}
M.~Ahmadizadeh, G.~Mosqueda, and A.~M. Reinhorn.
\newblock Compensation of actuator delay and dynamics for real-time hybrid
  structural simulation.
\newblock {\em Earthquake Engineering \& Structural Dynamics}, 37(1):21--42,
  2008.

\bibitem{Abbiati2018}
Giuseppe Abbiati, Vincenzo {La Salandra}, Oreste~S. Bursi, and Luca Caracoglia.
\newblock {A composite experimental dynamic substructuring method based on
  partitioned algorithms and localized Lagrange multipliers}.
\newblock {\em Mechanical Systems and Signal Processing}, 100:85--112, feb
  2018.

\bibitem{Abbiati2019a}
Giuseppe Abbiati, Igor Lanese, Enrico Cazzador, Oreste~S. Bursi, and Alberto
  Pavese.
\newblock {A computational framework for fast‐time hybrid simulation based on
  partitioned time integration and state‐space modeling}.
\newblock {\em Structural Control and Health Monitoring}, 26(10), oct 2019.

\bibitem{Craig1968}
Roy Craig and Mervyn Bampton.
\newblock {Coupling of Substructures for Dynamic Analyses}.
\newblock {\em AIAA Journal, American Institute of Aeronautics and
  Astronautics}, 6(7):1313--1319, 1968.

\bibitem{Patterson2013Aeo}
Michael~A. Patterson, Matthew Weinstein, and Anil~V. Rao.
\newblock An efficient overloaded method for computing derivatives of
  mathematical functions in matlab.
\newblock {\em ACM Trans. Math. Softw.}, 39(3):17:1--17:36, may 2013.

\bibitem{Quarteroni}
Alfio Quarteroni and Alberto Valli.
\newblock Theory and application of steklov-poincaré operators for
  boundary-value problems.
\newblock 01 1991.

\bibitem{Giagopulos2006}
D.~Giagopulos and S.~Natsiavas.
\newblock {Hybrid (numerical-experimental) modeling of complex structures with
  linear and nonlinear components}.
\newblock {\em Nonlinear Dynamics}, 47(1-3):193--217, dec 2006.

\bibitem{Tsotalou2017}
Maria Tsotalou, Dimitrios Giagopoulos, Vasilis Dertimanis, and Eleni~N. Chatzi.
\newblock {Model updating of a nonlinear experimental vehicle using
  substructuring and unscented Kalman filtering}.
\newblock {\em Proceedings of the 2nd International Conference on Uncertainty
  Quantification in Computational Sciences and Engineering}, pages 66--75,
  2017.

\bibitem{Pastor2012}
Miroslav Pastor, Michal Binda, and Tom{\'{a}}{\v{s}} Hararik.
\newblock {Modal Assurance Criterion}.
\newblock {\em Procedia Engineering}, 48:543--548, 2012.

\bibitem{Mucka2017}
P.~M{\'{u}}{\v{c}}ka.
\newblock {Simulated Road Profiles According to ISO 8608 in Vibration
  Analysis}.
\newblock {\em Journal of Testing and Evaluation}, 2017.

\bibitem{Rixen2004}
Daniel~J. Rixen.
\newblock {A dual Craig–Bampton method for dynamic substructuring}.
\newblock {\em Journal of Computational and Applied Mathematics},
  168(1-2):383--391, jul 2004.

\end{thebibliography}
\endgroup

\end{document}